\documentstyle[12pt]{article}
\begin{document}

\begin{center}
{\bf  QUANTUM MECHANICS AND THE COSMOLOGICAL CONSTANT}
\end{center}
\begin{center} {\bf L V Prokhorov}  \end{center}

\begin{center}
{V.A.Fock Institute of Physics \\
Sankt-Petersburg State University, Russia} \\
\end{center}

{\bf Abstract}. It is shown that in the model [3,4] of quantum mechanics
besides probability amplitudes, the Planck constant and the Fock space, the
cosmological constant also appear in the natural way. The Poisson brackets
are generalized for the case of kinetics.\\


\noindent
\section{\bf Introduction}

The problem of cosmological constant is one of the most difficult problems
of modern physics. It arises if one takes into consideration quantum
field theory (QFT). Indeed, the cosmological term $\Lambda g_{\mu\nu}$ in
the gravitational equations can be interpreted as presence of a kind of
vacuum energy density $\epsilon_{vac}^\Lambda$. According to the standard
QFT there is nonzero vacuum energy density $\epsilon_{vac}$, and it looks
natural to identify these two quantities.  Experiment gives [1]

\begin{equation}
\epsilon_{vac}^\Lambda \approx 2.5\cdot 10^{-67} {\rm cm}^{-4},
\end{equation}
But $\epsilon_{vac}$ is infinite, and simple estimation gives
$\epsilon_{vac} \sim L_0^{-4}$
where $L_0$ is some characteristic length. Taking  $L_0\sim 10^{-13}$ cm,
one gets

\begin{equation}
\epsilon_{vac} \sim 10^{52} {\rm cm}^{-4},
\end{equation}
discrepancy of 119 orders! Taking $L_0\sim l_{\rm P}=1.6\cdot 10^{-33}$ cm
worsens the situation.

There are
different approaches to the problem, e.g. (i) admit that $\Lambda$ is a new
fundamental constant of the theory, (ii) assume that there is a scalar field
with nonzero vacuum average and negative contribution to the vacuum energy,
and so on [2]. The first possibility means that there are two independent
scales in physics, while the second one assumes cancellation of enormous
numbers (fine tuning). Notice also that at the Planck scales the theory is
presumably supersymmetric. In this case the vacuum energy is zero, and the
$\Lambda$-term problem cannot be solved in the framework of high energy
physics (supersymmetric QFT).

It turns out, however, that the earlier proposed model of quantum mechanics
(QM) [3,4] as a phenomenon emerging at the Planck scales, gives natural
explanation of the second scale. The essential feature of the model ---
instability of the world: for an ordered set of classical oscillators in a
thermal bath QM (i.e. description by probability amplitudes) emerges in the
time interval of relaxation of its non-equilibrium state. The Planck
constant $h$ appears automatically. In this model all probability
amplitudes acquire a factor $\exp{(-\alpha t)}$ and decrease when time
$t\rightarrow\infty$. It implies non-stability of the Universe. All the
equations of motion for fields should be modified, masses increase,
$m^2 \rightarrow m^2+ \alpha^2$, so all the massless particles become
 massive. The cosmological term is nothing but half of the square of the
graviton mass ($m_g^2=\alpha^2=2\Lambda$). In particular, the model predicts
the same mass for photon ($m_{ph}^2 =2\Lambda$).

In Sec. 2 the concise presentation of the model [3,4] is given. It is
noteworthy that the model elucidates the nature of classical equations of
motion. For description of the non-equilibrium states evolution modified
Poisson brackets are proposed. The case of an oscillator is considered in
Sec. 3, and QFT --- in Sec. 4.  It is shown that as a result, all the
massless fields acquire masses $\alpha$.  For the gravitational field it
means appearance of $\Lambda= \alpha^2$. It allows to estimate the
temperature of the heat bath ($T\gg 10^{-29}$ K). In Sec. 5 the main results
of the article are summarized.

\section{\bf Quantum mechanics}

Probability amplitudes describe evolution of non-equilibrium
distributions of a harmonic oscillator in a thermal bath [3,4].  Let

\begin{equation}
G(q,p)=h^{-1}e^{-\beta H(q,p)},\quad H=\frac{\omega}{2}(p^2+q^2),\quad
\beta=1/kT,\quad h=2\pi/\beta \omega,
\end{equation}
be the Gibbs distribution ($k$ --- the Boltzmann constant). One may also use
complex variables $z=(q+ip)/\sqrt 2$, $\bar z$. This distribution introduces
a measure $\mu$ in phase space and defines a dimensional constant
$h$ (the area unit of the phase plane) which can be
identified with the Planck constant. The measure is

\begin{equation}
d\mu (q,p)=\frac{dq\wedge dp}{h}e^{-\beta \omega(p^2+q^2)/2}= d\mu (\bar
z,z)=\frac{d\bar z\wedge dz} {ih}e^{-\bar z z/\hbar}, \ \ H=\omega\bar z z,\
\ \hbar= \frac{h}{2\pi}.
\end{equation}
In transformation

\begin{equation}
\left(\begin{array}{c} \bar z \\ z  \end{array} \right)=\!
\hat {U}\left(\begin{array}{c} q \\ p \end{array} \right),\ \ \hat {U}=
\frac{1}{\sqrt 2}\left( \begin{array}{cc} 1&-i \\ 1&i \end{array} \right),
\end{equation}
matrix $\hat U$ is unitary ($\hat U^+\hat U=1$), and

\begin{equation}
H=\!\frac{\omega}{2} \left( \begin{array}{c} q \\ p \end{array}
\right) \left( \begin{array}{cc} 1&0 \\ 0&1 \end{array} \right) \left(
\begin{array}{c} q \\ p \end{array} \right)=\!
\frac{\omega}{2} \left( \begin{array}{c} \bar z \\ z \end{array}
\right)U^*U^+\left(\begin{array}{c} \bar z \\ z  \end{array} \right) =\!
\frac{\omega}{2} \left( \begin{array}{c} \bar z \\ z \end{array}
\right)\left( \begin{array}{cc} 0&1 \\ 1&0 \end{array} \right)
\left(\begin{array}{c} \bar z \\ z  \end{array} \right).
\end{equation}
The Poisson bracket in complex variables reads

\begin{equation}
\{f,g\}=
\left(\frac{\partial f}{\partial q}\frac {\partial g}{\partial
p}-\frac{\partial f}{\partial p}\frac{\partial g}{\partial
q}\right)\equiv \frac{\partial (f,g)}{\partial(q,p)}=
i\frac{\partial (f,g)}{\partial(\bar z,z)},
\end{equation}
i.e. transformation $(q,p)\rightarrow (\bar z,z)$ is not canonical one [5].
It is important that in case of complex variables one may use only one
equation of motion

\begin{equation}
\dot z =\{z,H\}=-i\omega z.
\end{equation}
Another equation $\dot {\bar z} =i\omega \bar z$ is obtained by complex
conjugation --- an essential feature of QM. Any other measure $\mu_p$,

$$d\mu_p=p(x)d\mu,\quad (x_1,x_2)=(q,p),\quad p(x)\geq 0,\quad\int
d\mu_p=1,$$ describes some non-equilibrium state [6].

Now, let's generate the non-equilibrium distributions.
Consider variation of $x$: an arbitrary variation $\delta x$ can be
given by the sum

\begin{equation}
\delta x=\delta x_{\perp}+\delta  x_{\parallel},
\end{equation}
where $\delta x_{\perp}$ by definition preserves the Gibbs distribution, i.e.

\begin{equation}
\delta H=\nabla H \delta x_{\perp}=0.
\end{equation}
Solutions of this equation are given by

\begin{equation}
\delta x_{\perp}=\hat J \nabla H \delta t,
\end{equation}
$\hat J$ being some antisymmetric matrix, and $t$ --- some parameter.
Identifying $t$ with time and taking $\hat J =
\left( \begin{array}{cc} 0&1 \\ -1&0 \end{array} \right)$, we observe that
(11) coincides with the classical equation of motion

\begin{equation}
\dot x =\{x,H\}.
\end{equation}

As for variations $\delta {\bf x}_{\parallel}$ --- they, evidently, destroy
equilibrium. Taking, e.g. small variations $z\rightarrow z+c$, we have
 $\bar z z\rightarrow (\bar z +\bar c)(z+c)$, and

\begin{equation}
d\mu(\bar z,z)\rightarrow d\mu_f(\bar z,z)=|f_c(z)|^2d\mu(\bar z,z), \ \
f_c(z)=e^{-\bar c z/\hbar}e^{-\bar c c/2\hbar}.
\end{equation}
Measure $\mu_f(\bar z,z)$ describes a non-equilibrium distribution which
tends to the equilibrium one with time ($|f_c|^2\rightarrow 1$).

There are two important points here.\\
1. To describe evolution of $\mu_f(\bar z,z)$  it is enough to take only
one equation of motion (see (8)) [3,4]

\begin{equation}
\dot f=\{f,H\}=-i\omega z\frac {df}{dz}.
\end{equation}
2. If the time of relaxation $t_r$ is large ($\omega^{-1}\ll t_r $), in the
time interval $0 < t \ll t_r$ one can use classical equation of motion
(14) (as for the "vacuum energy" $\hbar\omega/2$ here --- see below).

Measure (13) allows to introduce the Hilbert space of entire analytic
functions $f(z)$ of order $\rho \leq 2$. Indeed, from
linearity of Eq. (14) it follows that sum of its solutions $f_1(z),f_2(z)$
is also solution (the principle of superposition). Substituting
into (13) $f=f_1+f_2$ and integrating, one gets e.g. expressions like
$\int d\mu \bar f_1 f_2$ which can be identified with the scalar product of
vectors $f_1, f_2$.
Functions $f_c(z)$ for arbitrary small but finite $c$ form an overcomplete
basis in the Fock space [7] with the scalar product

\begin{equation}
(g(z),f(z))=\int d\mu(\bar z,z) \overline{g(z)}f(z).
\end{equation}

Thus, analytic functions $f(z)$ are \\
(i) dynamical variables, and\\
(ii) they can be identified with probability amplitudes (elements of the
Hilbert space).\\
State vectors $Z_n(z) = \tilde z^n/\sqrt {n!}\equiv \langle n|z\rangle$,
$\tilde z=z/\sqrt \hbar$, form an orthonormal basis in the Fock space
[3,4,7]:  $(Z_n, Z_m)=\delta_{nm}$,

\begin{equation}
f(z)=\sum_{n=0}^\infty c_n Z_n(z),
\end{equation}
and

\begin{equation}
||f||^2=(f,f)=\sum_{n=0}^\infty |c_n|^2 =1.
\end{equation}
It means that $|c_n|^2$ give a discrete probability distribution, and
$c_n$ also can be considered as dynamical variables and probability
amplitudes. From Eq. (14) one obtains

\begin{equation}
Z_n(t)=c_n e^{-i\omega nt}Z_n(0)=c_n(t)Z_n(0),\quad f(t)=\sum_{n=0}^\infty
c_n(t)Z_n(0)
\end{equation}
(here $Z_n(t)\equiv Z_n(z(t))$). The time dependence of $c_n$ is generated
by the Hamiltonian

\begin{equation}
H=\omega\sum_{n=1}^\infty n\bar c_n c_n,
\end{equation}
$\bar c_n, c_n$ being complex canonical variables with the Poisson bracket

\begin{equation}
\{f,g\}=i\sum_{n=1}^\infty \frac{\partial (f,g)}{\partial(\bar c_n,c_n)};
\end{equation}
then $\dot c_n= \{c_n,H\}=-i\omega nc_n$. We observe that effectively this
is a physical system with infinite number of degrees of freedom (as it
manifests itself in the time interval $0 < t \ll t_r$ --- influence of the
heat bath). It is easy to prove that if the state of an oscillator is
specified by a function $f$, then $|(g,f)|^2$ is probability for the
oscillator to be in the "state $g$". Normalized function $g(z)$ can be
considered as an element of some orthonormal basis. Expanding $f(z)$ over
elements of this basis one finds that $(g,f)$ can be identified with a
complex number $c_g$ (an analog of $c_n$), and then $|c_g|^2$ is the above
mentioned probability.

Operators appear in natural way. Multiplication of $\overline {g(z)}$ by
$\bar z$ changes $\overline {g(z)}$: $\bar z\overline {g(z)}=\overline
{g_1(z)}$, i.e. $\bar z$ can be considered as an operator $\hat{\bar z}$.
The result of multiplication $\overline {g(z)}$ by $z$ can be find from
(15) (use identity $z\exp{(-\bar z,z/\hbar)}=-\hbar d\exp{(-\bar
z,z/\hbar)}/d\bar z$ and integrate by parts): $\overline
{g_2(z)}=\hat z\overline {g(z)}=\hbar d\overline {g(z)}/d\bar z$. Operators
$\hat{\bar z},\hat z$ have correct commutation relation $[\hat z,\hat{\bar
z}]=\hbar$, and $[\hat q,\hat p]=[\hat {\bar z}+\hat z,i(\hat {\bar
z}-\hat z)]/2=i\hbar$. The Hamiltonian also becomes an operator, and
Eq. (6) gives:

\begin{equation}
\hat H=\frac{\omega }{2}(\hat {\bar z}\hat z+ \hat z\hat{\bar z})=
\hbar\omega (\bar z\frac{d}{d\bar z}+\frac{1}{2});
\end{equation}
here $\hat H$ contains the missing in (14) the zero point energy 
$\hbar\omega/2$. It is remarkable that the classical expression (6)
contains the recipe for ordering of $\hat {\bar z},\hat z$.

Eq. (14) after multiplication by $i\hbar$ and complex conjugation differs
from the Schroedinger equation

\begin{equation}
-i\hbar \dot {\overline f}(z)=\hat H \overline f (z)
\end{equation}
only by an additive constant $\hbar\omega/2$ in energy. Thus, we have
complete quantum theory of a harmonic oscillator.

In QM, stochastic by their nature wave functions satisfy the
deterministic Schroedinger equation. The present
model elucidates this phenomenon. Let's make another non-canonical
transformation: $(\bar z,z) \rightarrow (\varphi,r)$, $z=r\exp{(i\varphi)}$.
Then
$$\{f,g\}=i\frac{\partial (f,g)}{\partial(\bar z,z)}=i\frac{\partial
(f,g)}{\partial(\varphi,r)}\frac{\partial(\varphi,r)}{\partial(\bar
z,z)}=-\frac{1}{2r}\frac{\partial (f,g)}{\partial(\varphi,r)}.$$
Here $H=\omega r^2$, and equations of motion for $\varphi,r$ read

\begin{equation}
\dot \varphi=-\omega,\quad  \dot r=0
\end{equation}
(of course, they also follow from Eq. (8)). It means that in the case of
non-equilibrium distribution one canonical variable ($\varphi$) satisfies the
classical equation of motion (23), while the second one does not change
with time: being pure stochastic, $r$ is characterized by the Gauss
probability distribution $\exp{(-r^2/\hbar)}$. In the sum for energy $E_n=
\hbar \omega(n+1/2)$ the first term represents contribution of classical
motion: taking in (14) $f=z^n$, one gets $\omega_n=n\omega$. The second
term in the sum appears due to the thermal bath; it is the same for all
states. The energy is quantized ($\hbar\omega_n=n\hbar\omega$) due
to the classical motion.

The average values $\bar A$ of phase functions $A(\bar z,z)$ are defined by

\begin{equation}
\bar A \equiv\int d\mu(\bar z,z)
\overline{f(z)}\hat A(\hat {\bar z},\hat z)f(z)\equiv (f,\hat A f),
\end{equation}
and it is assumed that operators $\hat {\bar z},\hat z$ in $\hat A(\hat {\bar
z}, \hat z)$ are ordered.

Thus, in the framework of pure classical mechanics one gets description by
probability amplitudes $f(z)$ (quantum mechanics) because $|f(z)|^2\equiv
p(x)$ can be identified with probability density. This is true for
quasi-stationary non-equilibrium states. We have obtained all the principal
ingredients of QM:  the Planck constant $h$, the complex valued probability
amplitudes $f(z)$, the Fock space and the Schroedinger equation. Notice also
that the classical equations of motion for an oscillator follow, in fact,
from preservation of the Gibbs distribution.

\section{\bf Generalization of the Poisson brackets}

Eq. (22) gives approximate description of a quasi-stationary state. The
exact equation should describe kinetics, i.e. tending of a non-equilibrium
distribution to the equilibrium one. The simplest way to do this --- to
introduce a kind of friction into the oscillator equation of motion

\begin{equation}
\ddot q +\omega^2 q=0\rightarrow\rightarrow\ddot q +2\alpha \dot q +\omega^2
q =0,\quad \alpha >0.
\end{equation}
Then, evidently, $t_r\sim 1/\alpha$. One can introduce friction also into
the Hamiltonian equations (12) by substitution $\dot x\rightarrow\dot x+
\alpha x$. But all this looks as an ad hoc solution.
Description of non-equilibrium states refers to the fundamental
physics, and there should exist some more general approach in the framework
of the Hamiltonian mechanics. Problem: describe evolution of measure $\mu_f$.

The Hamiltonian mechanics of oscillators assumes existence of two
fundamental tensors.  The first is antisymmetric one, it defines symplectic
phase space and the Poisson bracket:

\begin{equation}
\omega_2(q,p)=\sum_{k=1}^n\omega^{-1}_k(q,p)dq_k\wedge dp_k,\quad
\{f,g\}=\sum_{k=1}^n\omega_k(q,p)\frac{\partial (f,g)}{\partial (q_k,p_k)}.
\end{equation}
The second one is symmetrical, and in case of $n$ oscillators it defines
their Hamiltonian

\begin{equation}
H(x)=\frac{1}{2}\sum_{k=1}^n\sum_{ij} h_{ij}^k x_i^k x_j^k,\ \ i,j=1,2.
\end{equation}
For a single oscillator the antisymmetric and symmetrical matrices are
correspondingly: $\hat J$, $\hat J _{ij} =\epsilon_{ij}=-\epsilon_{ji}$,
$\epsilon_{12}=1$, and $h_{ij}=\omega\delta_{ij}$. To describe kinetics one
has to take into consideration the variations $\delta x_{\parallel}$ in Eq.
(9). These variations are "parallel" to $\nabla H$, and it looks
reasonable to generalize the Poisson bracket

\begin{equation}
\{f,g\}\rightarrow\{f,g\}_\alpha \stackrel{def}{=}\{f,g\}_{-}
+ \{f,g\}_{+},
\end{equation}
where $\{f,g\}_{-}\equiv \{f,g\}$, and

\begin{equation}
\{f,g\}_{+} =-\alpha \sum_{k=1}^n \frac{1}{2}\sum_{ij}h^{ij}_k
\nabla_i^k f \nabla_j^k g,\ \ h^{il}h_{lj}=\delta^i_j.
\end{equation}
The second term in (28) takes into consideration variations changing the
Gibbs distribution. The equations of motion are

\begin{equation}
\dot f = \{f,H\}_{-}+ \{f,H\}_{+},
\end{equation}
and if $k=1$, then

\begin{equation}
\delta x^i_{\parallel}=-\alpha h^{ij}\nabla_j H\delta t.
\end{equation}

For the harmonic oscillator (3) we have
$h^{ij}=\frac{1}{\omega}\delta^{ij}$ and the equations of motion

$$\dot q=\omega p -\alpha q,\ \ \dot p=-\omega q -\alpha p$$

can be written in the form

\begin{equation}
Dq=\omega p,\qquad Dp=-\omega q, \qquad D\equiv \partial_t + \alpha.
\end{equation}

Applying operator $D$ to the first Eq. (32), one obtains

\begin{equation}
\ddot q +2\alpha \dot q +(\omega^2 +\alpha^2)q =0,\ \ \alpha\ll \omega.
\end{equation}

We see that $\omega^2$ gets an additive $\alpha^2$. Thus, for harmonic
oscillators the effect of friction can be taken into consideration by
substitution in the equations of motion

\begin{equation}
\partial_t \rightarrow D = \partial_t + \alpha.
\end{equation}
Notice that (i) now the least action principle is not valid, (ii) $\alpha$
is positive in (29), (33); the bracket (28) with $\alpha<0$ describes the
opposite process --- development of a non-equilibrium state.

\section{\bf The cosmological constant}

It is important that the rule (34) is true also for fields, because fields
are ordered sets of harmonic oscillators. Indeed, as is well known, the
Hamiltonian of countable set of oscillators in the 1D theory is (in proper
units) [3,4]

\begin{equation}
H=\int \limits_{-\pi}^{-\pi}d k \omega (k) a^*(k)a(k),\quad
\omega (k)^2=m^2 + 4 \sin^2\frac{k}{2} \approx m^2 + k^2,
\end{equation}
and, according to (33), masses change: $m^2 \rightarrow m^2 + \alpha^2$.
Notice however, that $\omega (k)$ in (35) has no direct relation to $\omega$
in (3) or (32): the latter refers to the primordial oscillators ($\omega$
manifests itself in $\hbar=1/\beta\omega$), while $\omega (k)$ characterize
collective excitations of their linear sets. In practice $\omega (k)\gg
\alpha$.

Thus, after substitution (34) all the massless fields become "massive". For
example, taking a scalar field $\phi$, we observe

\begin{equation}
(\Box -m^2)\phi=0 \rightarrow\rightarrow [\Box
-2\alpha\partial_t-(m^2+\alpha^2)]\phi=0 \quad (\Box=-\partial_t^2 +\Delta).
\end{equation}
In the atomic and subatomic physics the effect of $\alpha$ is
extremely small (if $t_r$ is of order of the life time of the Universe
$T_U\sim 10^{10} $ yr, then $\alpha \sim 10^{-33}$ eV).

It may seem that
this could have important consequences for gauge fields because mass terms
break gauge invariance. In reality the situation is more specific. \\

1) One cannot use substitution (34) in Lagrangians. Indeed, in this case
the Lagrangian $L=(\dot q^2/\omega -\omega q^2)/2$ for oscillator (3)
transforms into

\begin{equation}
L^{\prime}=L+\frac{\alpha}{\omega}\dot q q + \frac{\alpha^2}{2\omega}q^2,
\end{equation}
giving equation of motion

\begin{equation}
\ddot q +(\omega^2 -\alpha^2)q =0
\end{equation}
different from (33) (no friction term, "wrong" change of frequency).
Evidently, this is because the Hamiltonian variation principle fails in
this case (no Liouville theorem, kinetics). \\

2) Such a substitution breaks Lorentz invariance of the theory (the second
term in Eq. (36) is not an invariant). It looks reasonable because time and
the space coordinates lose their equivalence.\\

The absence of the Lagrangian formalism makes senseless the demand of "gauge
invariance". The reasonable strategy is:
in a gauge theory one excludes all the nonphysical variables [8]
and then makes substitution (34) in equations of motion. One can
also take the equations in an arbitrary gauge. E.g., for free
electromagnetic field in the Feynman gauge one obtains an analogue of Eq.
(36)

\begin{equation}
(\Box -2\alpha\partial_t-\alpha^2)A_\mu=0.
\end{equation}
The photon mass becomes important at extremely low energies $E\sim 10^{-33}$
eV.

The situation is different in gravity. The gravitational equations for the
metric tensor $g_{\mu\nu}$ also contain the second time derivatives. Thus,
one has

\begin{equation}
G_{\mu\nu}=R_{\mu\nu}-\frac{1}{2}Rg_{\mu\nu}\rightarrow  G_{\mu\nu} -
\alpha\partial_tg_{\mu\nu}- \frac{\alpha^2}{2}g_{\mu\nu}+...\; .
\end{equation}
The last term here is nothing but the cosmological term in the
Einstein-Hilbert equations ($\Lambda =\alpha^2/2$). Introduction of the
cosmological constant does not violate general covariance. It explains the
recently observed effect of remote galaxies acceleration [9]. Besides, the
constant $\alpha$ gives the life time of the Universe $T_U\sim t_r
\sim 1/\alpha$. From inequality $\alpha\ll\omega$ one can estimate from
below the temperature of the heat bath. We have: $\omega=T$ (in units $\hbar
= k=1$), i.e. $\alpha\ll T$. Taking $T_U\sim 1/\alpha$, we obtain $\alpha
\sim 10^{-29}$ K, so $T\gg 10^{-29}$ K.

Thus, in the present approach the cosmological term appears in the natural
way. The effective "vacuum energy density", it generates, is little to do
with the matter vacuum energy. Being the mass of graviton, $\alpha$
transforms the Newton potential into the Yukawa potential.
This gives visible explanation to the effect of acceleration of distant
galaxies [9].

\section{\bf Conclusions}

We summarize the main points of the article.

1. Description by probability amplitudes arises in the "non-equilibrium
Hamiltonian mechanics" of harmonic oscillators in a heat bath with large
time of relaxation (complex canonical variables, small deviations from the
equilibrium). Here everything is important:\\
(i) the harmonic oscillator: fields and strings are sets of harmonic
oscillators,\\
(ii) complex canonical variables $\bar z,z$: they lead to complex probability
amplitudes,\\
(iii) the heat bath: the sourse of chaos,\\
(iv) non-equilibrium distributions: description by probability amplitudes
appears in the classical physics only for non-equilibrium states,\\
(v) large time of relaxation ($t_r\gg \omega^{-1}$): in this case the system
can be considered as a quasi-stationary, and one can take $t_r =\infty$,\\
(vi) smallness of deviations from the equilibrium: arbitrary deviations can
radically change the Gibbs distribution and the classical equations of
motion.\\
Then one obtains a theory with the Planck constant $h$, the Fock space and
the Schroedinger equation. The harmony of order (the Schroedinger equation)
and chaos (probabilities), specific for quantum mechanics, is due to the
fact that there is a pair of canonical variables $\varphi, r$: one
of them ($\varphi$) is pure deterministic (no influence of the heat bath,
Eq. (23)), while another ($r$) is pure stochastic (it is characterized
by the Gibbs distribution).  This also elucidates the physics behind
the quantum oscillator energy spectrum $E_n=\hbar\omega(n+1/2)$: the
first term $\hbar\omega n$ is due to the classical motion ($\dot
z^n=-i\omega n z^n$), while the "vacuum energy" $\hbar\omega/2$ is
universal, it comes from $r$ due to the heat bath.

2. It looks quite satisfactory that the Hamiltonian mechanics admits natural
generalization to kinetics (Eqs. (28)-(30)).

3. Quantum description emerges only in the non-stable systems. Of course, it
does not invalidate the standard quantum mechanics: the latter can be
{\it defined} as a self-consistent theory, but in the framework of classical
mechanics it appears as kinetics.

4. Another positive feature of the approach --- necessity of non-stability.
Our Universe is not a stable system, and this agrees with the point 3.
Unexpectedly, there inevitably appears the cosmological constant $\Lambda$
which can be identified  with $\alpha^2/2$. It gives estimation $\alpha^{-1}
\sim 10^{10}$ yr.

5. All the oscillator (and field) excitations disappear with time (when
$t\rightarrow\infty$). It seems that our world is even less stable than that
what follows from the general relativity.

6. The Lagrangian formalism fails. As a result, the Lorentz invariance, the
gauge invariance and the general covariance also fail (or should be
reformulated).

7. Still, the problem of vacuum energy in QFT (Eq. (2)) is left out. Its
solution depends on the model of physical space. If one models 3D space as a
superstring network [10], then the superstring vacuum energy is zero.


\begin{thebibliography} {20}

\bibitem [1] {1} Eidelman S et al. 2004 {\it Phys. Lett.} {\bf B 592} 191

\bibitem [2] {2} Dolgov A D, Kawasaki M 2005 {\it Phys. Atom Nucl.} {\bf 68}
860

\bibitem [3] {3} Prokhorov L V 2004 quant-ph/0406079

\bibitem [4] {4} Prokhorov L V 2005 {\it Vestnik SPbGU} Ser. 4 Vyp. 4 3
(in Russian)

\bibitem [5] {5} Prokhorov L V, Shabanov S V 1997 {\it Hamiltonian mechanics
of gauge systems} (SPbGU University Press) (in Russian)

\bibitem [6] {6} Sinai Ya G 1996 {\it Introduction to ergodic theory}
(Moscow, FAZIS) P. 7 (in Russian)

\bibitem [7] {7} Bargmann V 1961 {\it Commun. Pure. Appl. Math.} {\bf 14}
187

\bibitem [8] {8} Prokhorov L V 1988 {\it Usp. Fiz. Nauk} {\bf 154} 299

\bibitem [9] {9} Riess A G et al. 1998 {\it Astron. J.} {\bf 116} 1009; 2004
{\it Astrophys. J.} {\bf 607} 665

\bibitem [10] {10} Prokhorov L V 2004 {\it Space as a net} (SPbGU, NIIKH)
(in Russian)

\end{thebibliography}
\end{document}